\documentclass[acmlarge, nonacm]{acmart}

\AtBeginDocument{%
  }

\acmConference[Conference acronym 'XX]{Make sure to enter the correct
  conference title from your rights confirmation emai}{June 03--05,
  2018}{Woodstock, NY}

\acmISBN{978-1-4503-XXXX-X/18/06}

\begin{document}

\title{VTutor: An Open-Source SDK for Generative AI-Powered Animated Pedagogical Agents with Multi-Media Output}

\author{Eason Chen}
\affiliation{%
  \institution{Carnegie Mellon University}
  \city{Pittsburgh}
  \state{PA}
  \country{USA}
}
\email{eason.tw.chen@gmail.com}

\author{Chenyu Lin}
\affiliation{%
  \institution{New York University}
  \city{New York}
  \state{NY}
  \country{USA}
}

\author{Xinyi Tang}
\affiliation{%
  \institution{Carnegie Mellon University}
  \city{Pittsburgh}
  \state{PA}
  \country{USA}
}

\author{Aprille Xi}
\affiliation{%
  \institution{Carnegie Mellon University}
  \city{Pittsburgh}
  \state{PA}
  \country{USA}
}

\author{Canwen Wang}
\affiliation{%
  \institution{Carnegie Mellon University}
  \city{Pittsburgh}
  \state{PA}
  \country{USA}
}

\author{Jionghao Lin}
\affiliation{%
  \institution{The University of Hong Kong}
  \city{Hong Kong}
  \country{Hong Kong}
}

\author{Kenneth R Koedinger}
\affiliation{%
  \institution{Carnegie Mellon University}
  \city{Pittsburgh}
  \state{PA}
  \country{USA}
}

\begin{abstract}
The rapid evolution of large language models (LLMs) has transformed human-computer interaction (HCI), but the interaction with LLMs is currently mainly focused on text-based interactions, while other multi-model approaches remain under-explored. This paper introduces VTutor, an open-source Software Development Kit (SDK) that combines generative AI with advanced animation technologies to create engaging, adaptable, and realistic APAs for human-AI multi-media interactions. VTutor leverages LLMs for real-time personalized feedback, advanced lip synchronization for natural speech alignment, and WebGL rendering for seamless web integration. Supporting various 2D and 3D character models, VTutor enables researchers and developers to design emotionally resonant, contextually adaptive learning agents. This toolkit enhances learner engagement, feedback receptivity, and human-AI interaction while promoting trustworthy AI principles in education. VTutor sets a new standard for next-generation APAs, offering an accessible, scalable solution for fostering meaningful and immersive human-AI interaction experiences. The VTutor project is open-sourced and welcomes community-driven contributions and showcases.

\end{abstract}




\begin{teaserfigure}
  \includegraphics[width=\textwidth]{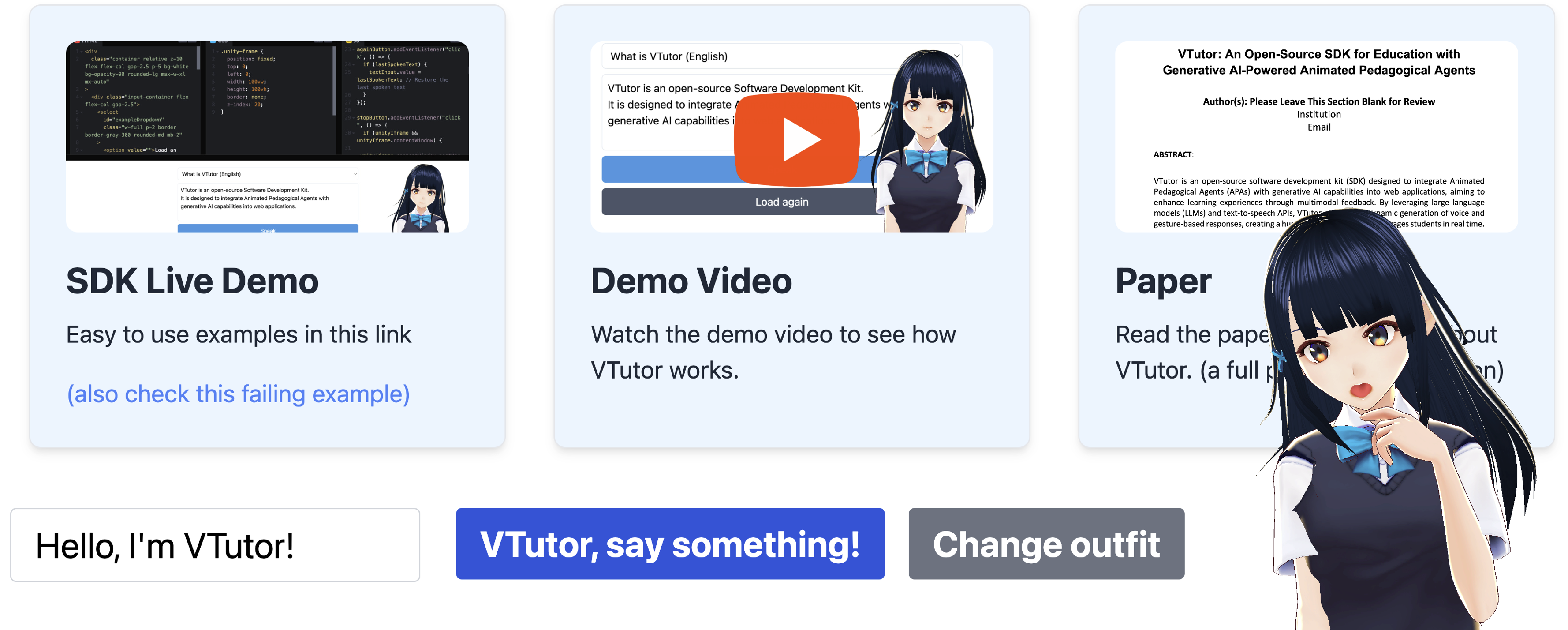}
  \caption{Screenshot of the VTutor demo website at \url{https://vtutor.tools}}
  \Description{Screenshot of the VTutor demo website at \url{https://vtutor.tools}}
  \label{fig:teaser}
\end{teaserfigure}

\maketitle

\section{Introduction}

The advent of large language models (LLMs) \cite{brown2020language} has greatly changed the research direction of Human-Computer Interaction (HCI) \cite{pang2025understandingllmificationchiunpacking} by enabling systems to generate coherent text, comprehend natural language, and adapt to a variety of contexts. These capabilities have paved the way for novel applications, from conversational agents \cite{chen2022effect} to task assistants \cite{mcnutt2023design} and tutoring systems \cite{wang2024large}. However, current implementations of LLM-driven systems are predominantly text-based, which limits the potential for creating more engaging and human-like interactions \cite{pang2025understandingllmificationchiunpacking}. While text-based interfaces are functional, they sometimes fall short in scenarios like giving feedback and learning companions scenarios \cite{Lin2024}.

Animated Pedagogical Agents (APAs) offer a compelling avenue to enhance and humanize these text-centric systems \cite{Lin2024}. Instead of engaging in purely text-based chat interactions in a chatbot interface or experiencing voice-only input and output akin to a phone call, APAs provide an experience akin to video chatting with a real person. APAs are on-screen characters that guide, tutor, and motivate learners. APAs demonstrated significant promise in educational contexts \cite{johnson2000animated}. By incorporating visual, vocal, and gestural cues, these agents can support adaptive or scripted roles \cite{Dai2022} and embody diverse appearances, from simplistic 2D avatars to intricate 3D humanoid figures capable of rich facial and body animations \cite{Domagk2010, Veletsianos2012, Makransky2018, Davis2021}. When well-designed, APAs can not only lower cognitive load but also provide emotional scaffolding, thereby improving learner motivation and retention \cite{Mayer2014}. However, current APA solutions are often constrained by limited voice realism, poor lip synchronization, and reliance on pre-scripted dialogue, hindering their ability to deliver dynamic, context-specific guidance at scale.

In this work, we bridge these gaps by introducing \textbf{VTutor}: an open-source Software Development Kit (SDK) that fuses generative AI techniques with cutting-edge animation technologies to elevate the realism, interactivity, and adaptability of APAs. Leveraging advanced lip synchronization engines \cite{githubGitHubHecomiuLipSync} and real-time WebGL rendering, VTutor enables seamless integration of naturalistic animated agents within web platforms. Crucially, it harnesses the adaptability of LLMs to timely and personalized dialogue for improved learner engagement. Through the use of VTutor, we aim to contribute to the evolving landscape of human-AI interactions by exploring how they can move beyond static text-based interfaces. VTutor combines generative AI with multi-media output, including synchronized text, speech, facial expressions, and animations, to deliver an engaging and immersive experience. By enabling more responsive and emotionally engaging interactions, we hope to support users in a variety of contexts, fostering guidance, motivation, and connection in a thoughtful and adaptive manner.
% Through the use of VTutor, we aim to reshape the landscape of human-AI interactions, transforming them from static text-based interfaces into responsive, emotionally resonant companions that can more effectively guide, motivate, and support users in various contexts.

\section{Related Works}
Animated Pedagogical Agents (APAs) are virtual characters designed to facilitate learning by providing guidance, feedback, and motivation to users. Initially conceptualized as tools to make learning more engaging, APAs have evolved significantly over the years. Their design is grounded in established theoretical frameworks, such as the persona effect, social agency theory, and embodied cognition theory, which emphasizes the importance of human-like interactions in fostering emotional connections, enhancing cognitive engagement, and improving knowledge retention \cite{Lester1997, Veletsianos2012, WangMultimedia}.

APAs have demonstrated their value across diverse educational domains, including electronics education \cite{Graesser2018} and language learning \cite{Roscoe2013}. Studies have shown that the use of APAs increases learner engagement and facilitates knowledge transfer by integrating multimodal cues, such as gestures, facial expressions, and speech, into the learning experience \cite{Makransky2018, Domagk2010}. By acting as social partners, APAs create interactive environments that encourage active learning and reduce cognitive load \cite{Mayer2014, Li2019}.

Despite their potential, existing APAs face critical limitations that hinder their widespread adoption and effectiveness. One major challenge is the lack of natural and emotionally resonant speech. Many APAs rely on rudimentary voice synthesizers, resulting in robotic and monotone outputs that fail to establish meaningful emotional connections with learners \cite{Dai2022}. Furthermore, lip synchronization technologies often fall short in aligning speech with mouth movements, leading to a jarring user experience. For example, Heeyo \cite{heeyoHeeyo}, an educational startup valued at \$20 million USD, employs APAs with only repetitive jaw-opening motions that do not correspond accurately to the spoken phonemes, reducing the sense of realism and immersion.

Another significant limitation lies in the adaptability of APA interactions. Many systems operate on script-based frameworks, predefining interactions and feedback that cannot dynamically adjust to the learner’s context or needs. On platforms like Cramify \cite{cramifyCramifyCram}, APAs are limited to pre-configured video-based interactions, offering no real-time customization of dialogue or feedback. This restricts the ability of APAs to provide personalized and responsive learning experiences, which are essential for addressing the unique challenges and preferences of individual learners.

In addition, a highly successful example of APAs with synchronized facial expressions and mouth movements is Duolingo’s video call feature \cite{duolingoVideoCall}, introduced in September 2024. This feature allows users to practice English through video calls with an APA named Lilly, sparking widespread discussion online. However, Duolingo didn't open-source the underlying technologies, and the complexity of building APAs brings significant technical challenges for developers and researchers who wish to implement their own APAs for experiments and applications.

We developed VTutor to address the aforementioned challenges. Unlike conventional script-based APAs, VTutor can use LLMs to generate personalized, context-specific guidance and feedback in real-time with matched facial expressions, creating a more engaging, human-like interacting experience. By leveraging these innovations, VTutor aims to redefine the role of APAs in education, making them more effective and accessible across diverse interaction environments.

\section{System Implementations}

\subsection{Text-to-Speech (TTS) Voice Generation}

VTutor seamlessly integrates text, voice, and animation into a cohesive multi-media output framework, enhancing the quality of learner interaction. To use VTutor, developers need to prepare audio by TTS in a \texttt{.wav} format to integrate with VTutor. Specifically, they can use TTS services such as OpenAI, Azure, Google Cloud Platform, and others to generate voice outputs for interaction with VTutor. Additionally, users can prepare real-time audio streams to enable more fine-grained and responsive interactions. In our demo website, we currently use Azure's speech engine as an example.

\subsection{Lip Synchronization (LipSync)}

For the lip synchronization (LipSync) component of VTutor, we leverage \textbf{uLipSync} \cite{githubGitHubHecomiuLipSync}, an open-source tool designed for real-time and preprocessed lip-syncing in Unity. uLipSync works by analyzing audio waveforms in real time using Mel-Frequency Cepstrum Coefficients (MFCC) to extract human voice characteristics. These coefficients represent phonemes (e.g., "a," "e", "i," "o", "u") by capturing vocal tract features. The tool then maps these phonemes to pre-configured blend shapes in a 3D model's \texttt{SkinnedMeshRenderer}, dynamically adjusting the model's mouth movements to match the audio input. This enables accurate synchronization between speech and visual expressions, significantly enhancing the realism and immersion of the animated pedagogical agents.

By integrating uLipSync into our framework, we allow developers to input either live microphone audio or pre-recorded audio clips to drive lip synchronization. The tool also supports pre-baked lip-sync data for high-performance scenarios, such as browser-based WebGL deployments. Moreover, VTutor uses customizable profiles to calibrate phonemes, ensuring compatibility with diverse voice types and accents. This approach resolves common issues with traditional animated pedagogical agents, such as unnatural or repetitive jaw movements, and ensures seamless alignment between spoken content and visual expressions, improving the learner's interaction experience.

\subsection{VTutor Character Model Selection}
VTutor incorporates both 2D and 3D character models developed within Unity, ensuring seamless compatibility with TTS and LipSync. This design choice underscores the framework's versatility, enabling users to customize their APAs by importing their preferred character models, including anime-style avatars. By supporting both 2D and 3D assets, VTutor provides unparalleled flexibility for creating tailored and engaging educational agents.

The TTS and lip-sync features are designed to be universally adaptable, allowing users to effortlessly integrate models sourced from popular online repositories and 3D model stores \footnote{such as \url{https://hub.vroid.com} and \url{https://sketchfab.com/search?category=people&features=animated&type=models}}. Thanks to Unity's standardized skeletal system, pre-defined gestures can be applied seamlessly across various character models, reducing the need for extensive adjustments and streamlining the customization process. That is, developers can easily import a preferred character model and compile a new VTutor instance to use it with the SDK on their website.

To showcase the framework's interoperability, we utilized two prominent open-source modeling platforms: Live2D for 2D character representation and VRoid Studio for 3D avatars. Both platforms offer models with foundational lip animation capabilities, which VTutor dynamically synchronizes with generated speech via its lip-sync technology. This ensures that APAs deliver content in a visually realistic and engaging manner, enhancing the learner experience.

This robust adaptability positions VTutor as a practical and user-friendly solution for educators and developers, expanding its utility across a diverse range of educational and creative applications. By simplifying the integration and customization process, VTutor empowers users to create personalized animated agents that elevate the quality and interactivity of learning environments.

\begin{figure}[h!]
    \centering
    \includegraphics[width=1\linewidth]{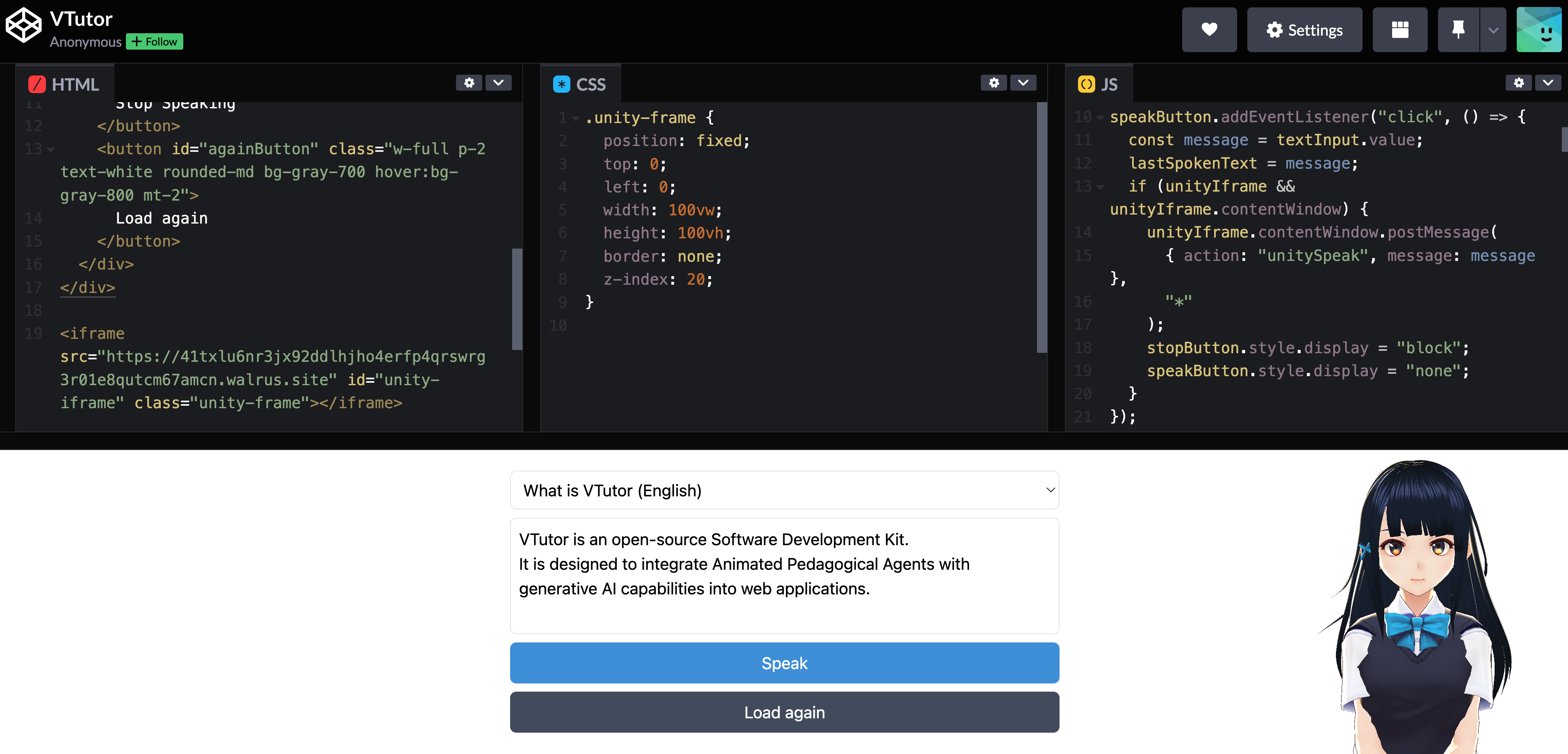}
    \caption{Screenshot of the VTutor SDK demo at \url{https://codepen.io/AnonymousForReview/pen/xxvaxJg}, the few lines of codes in the image are what developers need to do to add VTutor to their own website.}
    \label{fig:enter-label}
\end{figure}

\subsection{API Communication}
The primary application scenario for VTutor is in the web interface. To facilitate communication between the web interface and VTutor running in Unity WebGL, we leverage the \texttt{instance.SendMessage} method to transmit audio files in the WAV format. Once VTutor receives the WAV audio input, it processes the file using uLipSync and synchronizes the APA body and lip animations with the audio playback in real time.

\subsection{Software Development Kit (SDK) Implementations}
To enable seamless integration of VTutor into existing websites, we encapsulated the entire VTutor system, including its unity instance, web interface, and communication commands, within an \texttt{iframe}. This approach allows any website to embed VTutor quickly with minimal effort. As demonstrated in our example codes \footnote{\url{https://codepen.io/AnonymousForReview/pen/xxvaxJg}}, VTutor's iframe-based integration simplifies deployment with just a few line of codes. 

Furthermore, we are developing example codes and React SDK to facilitate smooth integration into React-based websites. This SDK provides developers with a streamlined workflow, enabling them to embed VTutor's features effortlessly while maintaining compatibility with modern web development practices. The React frontend example of VTutor is also open-sourced at \url{https://github.com/VTutorTools/vtutor-vercel-app}.

\section{Implications}

The VTutor project introduces a powerful framework that combines generative AI and advanced animation technologies to create emotionally engaging and highly adaptable Animated Pedagogical Agents (APAs). This innovation has significant implications for enhancing human-AI interaction and educational technologies, as outlined below.

\subsection{Advancing Animated Pedagogical Agents}
VTutor redefines the role of APAs by addressing key limitations in existing research and development. By leveraging open-source, large language models (LLMs), real-time text-to-speech synthesis, precise lip synchronization, VTutor enables:

\begin{itemize}
    \item \textbf{Personalized Learning}: Adaptive, context-specific feedback tailored to individual learners, improving engagement and knowledge retention.
    \item \textbf{Immersive Interactions}: Emotionally expressive agents with lifelike gestures, speech, and facial expressions create a strong sense of presence and relatability.
    \item \textbf{Accessible Development}: A developer-friendly SDK that allows seamless integration of APAs into web platforms, reducing the barrier to APAs adoption and research.
\end{itemize}

These contributions make VTutor a scalable and impactful solution for advancing APA technologies across domains.

\subsection{Potential Applications}
The versatility of VTutor's design opens up opportunities for impactful applications across various fields:
\begin{itemize}
    \item \textbf{Education}: Interactive tutoring systems that adapt to learners' progress, providing tailored guidance in STEM, language learning, and other domains.
    \item \textbf{Corporate Training}: Employee training programs that use personalized agents to deliver real-time feedback and improve knowledge retention.
    \item \textbf{Healthcare}: Virtual assistants for mental health support or patient education, offering empathetic, context-aware interactions.
    \item \textbf{Entertainment}: Engaging storytelling agents or interactive gaming characters that respond dynamically to player inputs.
\end{itemize}
By enabling these applications, VTutor demonstrates its potential to reshape how humans interact with AI in educational, professional, and recreational settings.

\section{Conclusion}

The VTutor project presents a groundbreaking framework that combines the capabilities of generative AI with advanced animation technologies to redefine the potential of Animated Pedagogical Agents (APAs) in education. By addressing the limitations of traditional APAs, such as scripted dialogues, limited adaptability, and lack of realism, VTutor introduces a versatile and scalable Software Development Kit (SDK) that enables developers and educators to create personalized, interactive, and emotionally engaging learning experiences. Through its integration of large language models (LLMs), cutting-edge lip synchronization engines, and seamless WebGL rendering, VTutor offers a powerful toolkit for building human-like pedagogical agents that foster learner engagement, enhance feedback receptivity, and establish trustworthy AI interactions.

Crucially, VTutor is an open-source project, emphasizing accessibility, transparency, and community-driven innovation. We welcome contributions from developers, educators, and researchers around the world to further improve the SDK. Whether it involves integrating new character models, enhancing system features, developing additional examples, or suggesting innovative improvements, we invite pull requests and collaborative efforts. By fostering an open and inclusive community, VTutor aims to accelerate the adoption and advancement of generative AI-powered APAs, ultimately shaping the future of educational technology.

The source code, examples, and documentation for VTutor are available at \url{https://github.com/VTutorTools}, and we encourage the broader community to join us in refining and expanding this transformative toolkit with feedback, pull requests, and example showcases. Together, we can unlock the full potential of human-AI interaction, paving the way for a more engaging, inclusive, and effective interaction experience.

% \begin{acks}
% This paper is partly supported by the Generative AI + Education Tools: Seed Grant Program at Carnegie Mellon University
% \end{acks}

\bibliographystyle{ACM-Reference-Format}
\bibliography{reference}

\end{document}